\documentclass[
twocolumn,
letterpaper,
runinaddress,
preprintnumbers,
nofootinbib,
 amsmath,amssymb,amsthm,
 aps,
prd,
]{revtex4-1}

\usepackage{parskip}
\usepackage{lipsum}
\usepackage{dcolumn}
\usepackage{bm}
\usepackage{tikz}
\usetikzlibrary{matrix}
\usepackage{slashed}
\usepackage{fontenc}
\usepackage{graphicx,hyperref,color,xcolor}
\usepackage[scr=rsfso,cal=zapfc,frak=euler,bb=ams]{mathalfa}

\definecolor{awesome}{rgb}{1.0, 0.13, 0.32}
\definecolor{electricblue}{rgb}{0.03, 0.57, 0.82}
\definecolor{guppiegreen}{rgb}{0.0, 0.88, 0.4}
\definecolor{blue-violet}{rgb}{0.54, 0.17, 0.89}

\makeatletter
\def\p@figure{\color{blue-violet}}
\def\p@equation{\color{blue}}
\makeatother

\hypersetup{
    bookmarks=true,         
    unicode=false,          
    pdftoolbar=true,        
    pdfmenubar=true,        
    pdffitwindow=false,     
    pdfstartview={FitH},    
    pdftitle={On the temperature of the quantum black hole},    
    pdfauthor={Abram Akal},     
    pdfcreator={Abram Akal},   
    pdfnewwindow=true,      
    colorlinks=true,       
    linkcolor=blue-violet,          
    citecolor=blue,        
    filecolor=blue,      
    urlcolor=blue          
}

\newcommand{\specChar}[1]{\mathsf{#1}}

\usepackage{titlesec}
\titleformat*{\section}{\bfseries}
\titleformat*{\subsection}{\itshape}
\titleformat*{\paragraph}{\bfseries\itshape}

\begin{document}

\title{On the temperature of the quantum black hole}
\author{Abram Akal}
\email[E-mail: ]{ibraakal@gmail.com}
\affiliation{Institute for Theoretical Physics\\
Utrecht University\\
3584 CC Utrecht, Netherlands}
\thanks{Author's affiliation when he started writing up this paper.}
\date{\today}

\begin{abstract}
A nontrivial peculiarity of general relativity is that when the horizon region of black holes is rendered harmless, the exterior doubles, resulting in a causally disconnected parallel universe. This intricacy plays a central role in ’t Hooft’s unitarity arguments, emphasising an exact identification between the physical universe and its duplicate on the other side of the horizon. However, it leads to another tension in the form of a factor of two correction in Hawking’s temperature. This discrepancy is concerning because the Rindler temperature is universal and complies with the Bekenstein-Hawking entropy. We demonstrate that the mismatch in the Boltzmann factor gets fixed if the state that forms the corresponding density matrix adopts a generalised thermofield double structure. That leaves room for some interesting discussion.
\end{abstract}

\maketitle

\section*{Motivation}
\label{sec:mot}
Following the seminal insights of Bekenstein \cite{Bekenstein:1972tm, Bekenstein:1973ur} and Hawking \cite{Hawking:1974rv, Hawking:1975vcx, Hawking:1976ra}, the quest for a consistent reconciliation of the principles of quantum mechanics and general relativity has gained ever more significance. Interestingly, there is an intrinsic feature of relativity that enables the consistent application of quantum theory to gravitational objects such as black holes. This is nothing but the equivalence principle. It says that any homogeneous gravitational field can be viewed as being generated by an acceleration relative to an inertial frame in which gravity is absent. From the working rules in the reference system, one would then be able to derive the rules for the system exposed to the gravitational field. 

In quantum theory, the Hamiltonian describes the time evolution in Hilbert space. However, the latter can only be well defined if the background is static. When the condition of time independence is mildly violated, a Hilbert space may still be definable if principles like locality and causality apply. This relaxation will be of central relevance for understanding the quantum aspects of black holes.

\begin{figure}[h]
    \centering
    \includegraphics[width=0.5\linewidth]{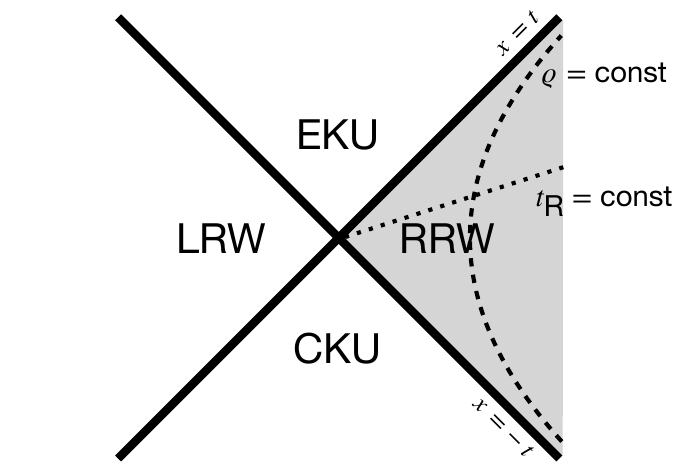}
    \caption{Shown are the shaded right Rindler wedge ($\specChar{RRW}$), i.e. Rindler space, left Rindler wedge ($\specChar{LRW}$), expanding Kasner universe ($\specChar{EKU}$), and contracting Kasner universe ($\specChar{CKU}$). The thick line corresponds to the Rindler horizon. Constant $\varrho$ (dashed) and $t_\text{R}$ (dotted) lines are also depicted. Note that the entire Rindler space is causally connected.}
    \label{fig:rindler-space}
\end{figure}

A gravitational field is generated and one-to-one mapping to an inertial frame if an observer is uniformly boosted through Minkowski space. This is the well-known Rindler space \cite{Fulling:1972md, Davies:1974th}. As the field theoretic quantum vacuum is Lorentz invariant, the observer, who is exposed to the generated field, experiences a thermal Gibbs state represented by a density matrix of the form
\begin{equation}
    \rho_\text{R} = \frac{e^{-\beta_\text{R} H_\text{R}}}{Z},
\label{eq:rindler-denisty-matrix}
\end{equation}
where $Z \equiv Z(\beta_\text{R})$ is the canonical partition function. The thermal character of $\rho_\text{R}$ can be traced back to the formation of the Rindler horizon, cf.~figure~\ref{fig:rindler-space}.

The Boltzmann factor in \eqref{eq:rindler-denisty-matrix} depends on the dimensionless inverse Rindler temperature,
\begin{equation}
    \beta_\text{R} = 2 \pi.
\label{eq:rindler-temperature}
\end{equation}
Importantly, recall that the appearance of the vacuum as a canonical ensemble, reflected through \eqref{eq:rindler-denisty-matrix} and the value in \eqref{eq:rindler-temperature}, does not depend on the details of the relativistic quantum field theory.
The Rindler Hamiltonian $H_\text{R}$ is expressed in terms of the usual Hamiltonian density in Minkowski space, weighted with the inverse of the observer's proper acceleration $a$ at some proper distance $1/a$ from the Rindler horizon. The proper temperature $T_\text{U} = a/\beta_\text{R} = a T_\text{R}$ is known as the Fulling-Davies-Unruh temperature \cite{Fulling:1972md,Davies:1974th,Unruh:1976db}, where $T_\text{U}$ increases as one approaches the Rindler horizon.

Now, if $a$ is taken to be the surface gravity of the Schwarzschild solution, the Unruh temperature $T_\text{U}$ equals Hawking's result for the temperature of black hole radiation, $T_\text{H} = 1/ \beta_\text{H} = 1/(8 \pi G M)$ \cite{Hawking:1975vcx}.
From the first law of thermodynamics $dE = T dS$,
the Bekenstein-Hawking entropy $S_\text{BH} = A/(4 G)$ follows straightforwardly.
In this context, the black hole entropy derived from $\rho_\text{R}$ has a coarse-grained thermodynamic character.

\begin{figure}[h!]
    \centering
    \includegraphics[width=0.5\linewidth]{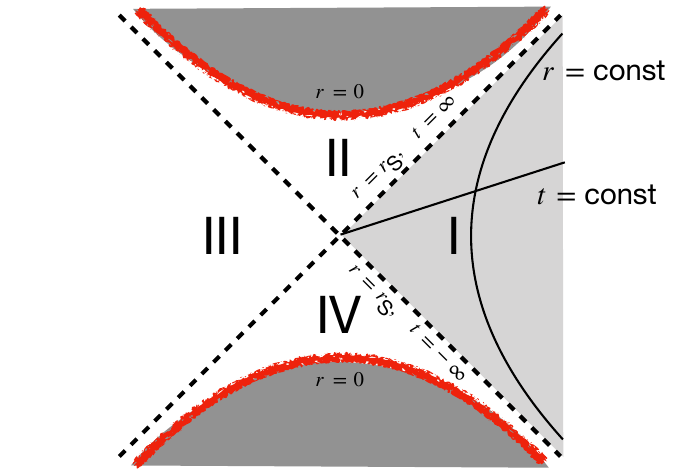}
    \caption{Maximally extended Fronsdal-Kruskal-Szekeres diagram of the Schwarzschild black hole solution. The coordinates are well behaved everywhere outside the curvature singularity shown as a thick, fuzzy (red) curve. In particular, the past and future horizon regions are non-singular. Shown are all four causally disconnected regions $\specChar{I}$ - $\specChar{IV}$. The physical universe corresponds to the shaded exterior region $\specChar{I}$. The darkened area in the interior is blocked due to the curvature singularity at $r=0$. The dashed lines represent the past and future horizons at $r=r_\text{S}$. Constant $t$ and $r$ lines are depicted as well.}
    \label{fig:schwarzschild-ks}
\end{figure}

This entropy can be equally viewed as a fine-grained entanglement entropy. To see this, one purifies the right Rindler wedge ($\specChar{RRW}$), i.e., the Rindler space, by maximally entangling its states with the states of the left Rindler wedge ($\specChar{LRW}$), resulting in a thermofield double state and a pure density matrix. By tracing out the degrees of freedom assigned to $\specChar{LRW}$, the density matrix associated with the Rindler space is reduced to $\rho_\text{R}$. This procedure reproduces $T_\text{H}$ \cite{Israel:1976ur}.

The two equivalent formulations concerning the nature of the entropy constitute an instantiation of the mixture equivalence principle. To be more precise, consider the proper mixed density matrix
\begin{equation}
    \rho^\text{prop} = p |a\rangle \langle a| + (1-p)|b\rangle \langle b|
\label{eq:proper-density-matrix}
\end{equation}
and the improper entangled state;
\begin{equation}
  |\psi\rangle_\text{improp} = \sqrt{p} |a,a\rangle + \sqrt{1-p}|b,b\rangle.
\label{eq:improper-state}
\end{equation}
Performing a partial trace over the space $\mathcal{H}_a$ yields the same reduced density matrix $\rho_b$ for both states, i.e.,
\begin{equation}
    \mathrm{tr}_a(\rho^\text{improp}) \equiv \rho_b^\text{improp} = \rho_b^\text{prop} \equiv \mathrm{tr}_a(\rho^\text{prop}).
\end{equation}
According to the mixture equivalence principle, the proper and improper mixed states with an identical density matrix are empirically indistinguishable. In other words, they would result in the same expectation value for a given operator $\mathcal{O}$ representing an observable, i.e.,
\begin{equation}
    \mathrm{tr}(\mathcal{O} \rho_b^\text{improp}) = \langle \mathcal{O}\rangle = \mathrm{tr}(\mathcal{O} \rho_b^\text{prop}).
\label{eq:oev-proper-improper}
\end{equation}
Quantum mechanically, nothing is confusing with this principle. However, the picture changes when gravity is taken into account. In the present context, this is demonstrated through the proper density matrix \eqref{eq:rindler-denisty-matrix} and the improper thermofield double state 
\begin{equation}
 | \Psi \rangle_\text{LRW,RRW} = \frac{1}{\sqrt{Z}} \sum_n e^{- \beta_\text{R} E_n / 2}\ | n \rangle_\text{LRW} \otimes | n \rangle_\text{RRW}.
\label{eq:main-rindler-tfd}
\end{equation}
Thus, semiclassical derivations of the radiation spectrum seem to lead to an improper state \cite{Fedida:2025fym}. Accordingly, a severe internal ambiguity arises with respect to the nature of the physical spacetime geometry underlying the proper state.

Of course, one may argue that this is a puzzle within the incomplete semiclassical approach. Nevertheless, we think that consistency in the physical exterior complying with the principles of both general relativity and quantum theory will not imply a dramatic change in the causal structure of the underlying geometry. In the quantum gravitational regime, the definition of proper and improper states may therefore be closely tied to a unique causal boundedness. Indeed, as we discuss shortly, no distortion of that sort will be implied through the semiclassical limit. In this sense, the quantum black hole prevents situations where the mixture equivalence principle is violated or enforces an ambivalence in the causal boundedness of the physical spacetime geometry. It is somewhat similar to the fact that the classical equivalence principle is kept intact throughout the entirety of the description.

Due to the character of $\rho_\text{R}$, the expectation values of the operators acting in the Rindler space would not comply with the preservation of information. To restore unitarity, 't Hooft argues that states of $\specChar{LRW}$ should be identified as the bra states of the $\specChar{RRW}$ states \cite{tHooft:1983kru}. However, a transition from the usual entangled state to a formulation based on the density matrix of the Rindler space implies a factor of two correction in the corresponding temperature \cite{Unruh:1976db}.

It should be pointed out that both Rindler wedges describe independent universes, so there is no principled justification for identifying the states of these two causally disconnected regions. Even though this argument seems plausible in Rindler space, for the quantum black hole, a more severe constraint could indeed alter the situation and cancel any ambiguity. Hence, assuming unitarity in the physical exterior may enforce an identification between the states of region $\specChar{I}$ and $\specChar{III}$ in the maximal extension of the Schwarzschild solution, cf.~figure \ref{fig:schwarzschild-ks}. In fact, the responsible Fronsdal-Kruskal-Szekeres \cite{Fronsdal:1959zza, Kruskal:1959vx, Szekeres:1960gm} coordinates behave well outside, around, and past the future and past horizon regions up to the curvature singularity in the interior. Thus, a correction in the temperature of the quantum black hole seems to be required \cite{tHooft:1983kru,tHooft:2023ggx}.

We take on the identification philosophy, as we currently do not see a physical reason to miss it if only the principles of quantum theory and general relativity are invoked without going beyond with more intricate and less verifiable frameworks. However, we also raise concerns regarding the corrections in the thermodynamic values, which, eventually, are quite robust predictions even without string theory or AdS/CFT.

There is at least one more puzzle that we seem to be uncomfortable with. The identification prescription described above yields a density matrix for the physical exterior $\specChar{I}$. However, this density matrix cannot be linked to the thermal Gibbs state via some appropriate map. In other words, it does not incorporate a semiclassical limit. From this point of view, it lacks a clear operational meaning. In particular, one would like to know the procedure for taking the limit in which $\rho_\text{R}$ reappears. We will see that this comes naturally in this paper. 

Here, we wish to ask what kind of quantum state of the physical region $\specChar{I}$ would comply with the identification prescription and leave the temperature and entropy of the quantum black hole unchanged. We point out that some sort of generalised thermofield double state interestingly achieves this. Our intention is not to propose the ultimate density matrix of the physical exterior. We rather demonstrate that an additional (protected) quantum structure renders the identification picture compatible with the established thermodynamic values.

In the following two sections, we introduce the generalised state and discuss how it fits within the outlined picture. In the last section, we conclude.

\section*{Thermofield double universe}
\label{sec:tfd-universe}
We shall first comment on the near-horizon thermodynamics from the perspective of a fiducial observer. The relation between the Schwarzschild time $t$ and the observer's proper time $\tau$ at a given Schwarzschild radius is given by
\begin{equation}
    \frac{d\tau}{dt} = \sqrt{g_{00}} = \sqrt{1 - \frac{r_\text{S}}{r}}.
\label{eq:rel-fido-proper-time-between-global-time}
\end{equation}
The near-horizon dynamics can effectively be described in the form of a fictitious thermal timelike membrane (or stretched horizon) when viewed from the perspective of the observer being at rest in the asymptotic rest frame and time $t$ of the black hole \cite{Price1986MembraneVO}. Ideally, the observer's proper time $\tau$ shall agree with the Schwarzschild time $t$, which, of course, strictly applies in the infinite spatial distance as can be deduced from \eqref{eq:rel-fido-proper-time-between-global-time}. We may position the stretched horizon slightly outside the true horizon at some finite proper distance $\Delta_\text{sh}$, ensuring that a freely falling observer does not witness any drama.

The universal properties of the stretched horizon can be made evident in the near-horizon region. Its local geometry for the Schwarzschild black hole is described by the Rindler space. The Rindler Hamiltonian $H_\text{R}$ is by definition conjugate to the Rindler time $t_\text{R}$, i.e., $[H_\text{R}, t_\text{R}] = i$. The eigenvalue of the Hamiltonian is the dimensionless Rindler energy $E_\text{R}$. Considering $E_\text{R}$ as a function of the mass $M$ of a sufficiently massive black hole, the relation between both quantities is shown to be of the following form \cite{Susskind:1993ws}
\begin{equation}
    E_\text{R} = 2 G M^2.
\label{eq:rindler-energy}
\end{equation}
We see that restoring the entropy of the black hole in \eqref{eq:rindler-energy} yields the first law of thermodynamics, i.e.
\begin{equation}
    dE_\text{R} = dS /\beta_R,
\label{eq:nh-1st-law}
\end{equation}
indicating the universal thermal character of the stretched horizon. Of course, to obtain the proper temperature seen by the distant Schwarzschild observer, the Rindler temperature $1/\beta_\text{R}$ has to be redshifted, which gives rise to the standard value $1/\beta_\text{T}$ of the Hawking temperature.

We may therefore expect that, in any consistent description in the physical exterior, correlations will effectively decouple from the microscopic dynamics of the stretched horizon and be liberated if kept sufficiently far away from the black hole. Equivalently, one may argue that local operations taking place in the far distance will be blind to the complex near-horizon dynamics. We further call to mind that it is appropriate to assume
\begin{equation}
    \Delta_\text{sh} \gtrsim \sqrt{G} \equiv l_\text{P}.
\label{eq:min-proper-distance-sh}
\end{equation}
In what follows, we write $\beta \equiv \beta_\text{R}$ for the inverse of the Rindler temperature, and bear in mind that for the proper temperature of the distant fiducial observer the gravitational red shift effects need to be restored by the simple rescaling $\beta \rightarrow \beta / \kappa$, where $\kappa$ denotes the surface gravity for the Schwarzschild solution.

Now, let us discuss the density matrix of the physical exterior of the black hole. As mentioned, the identification shall be made between the regions $\specChar{I}$ and $\specChar{III}$ of the maximally extended spacetime. In doing so, the states of the region $\specChar{III}$ are treated as the bra states ${}_\specChar{I}\langle n |$ of the states $| n \rangle_\specChar{I}$ that occupy the region $\specChar{I}$. The density matrix of the system in the region $\specChar{I}$ is proposed to be \cite{tHooft:1983kru}
\begin{equation}
    \rho^\text{g}_\specChar{I} = \frac{1}{Z} \sum_n e^{- \pi E_n}\ | n \rangle_\specChar{I}\ {}_\specChar{I}\langle n |,
\label{eq:rho-g-I}
\end{equation}
leading to the mentioned factor of two correction, i.e., $T = 2 T_\text{H}$ and $S = S_\text{BH} / 2$. Hence, with respect to the value of the Boltzmann factor, the matrix \eqref{eq:rho-g-I} deviates from the thermal Gibbs state represented by
\begin{equation}
    \rho_\specChar{I}^\text{semi} = \frac{1}{Z} \sum_n e^{- \beta E_n}\ | n \rangle_\specChar{I}\ {}_\specChar{I}\langle n |.
\label{eq:thermal-density-matrix}
\end{equation}
We wish to stress that \eqref{eq:rho-g-I} appears to be somewhat inappropriate since it is maximally mixed. This necessarily must not be that problematic, as one could argue that the black hole is viewed to be at a specific stage during its evolution. However, it is not clear how the matrix $\rho^\text{g}_\specChar{I}$ could be purified or whether purity is anticipated at all.

The situation is clearly different for the improper Hartle-Hawking or thermofield double state \eqref{eq:main-rindler-tfd} for which purification is performed through the region $\specChar{III}$. Recall that from the point of view of the fiducial observer sitting in $\specChar{I}$, the future and past interiors, i.e., $\specChar{II}$ and $\specChar{IV}$, respectively, are considered not physical.

In light of our discussion, let us then introduce the following simplified symmetric state (i.e., having equal energy eigenvalues for both parties) that is supposed to describe the physical exterior $\specChar{I}$,
\begin{equation}
    | \Psi \rangle^\text{qbh}_\specChar{I} = \frac{1}{\sqrt{Z}} \sum_n e^{- \beta E_n / 2}\ | n \rangle_\specChar{sh} \otimes | n \rangle_\specChar{out}.
\label{eq:qbh-state}
\end{equation}
As is the case for the improper state \eqref{eq:main-rindler-tfd}, this state is also invariant under the original modular Hamiltonian. However, note that a slight modification is needed as the left physical Hamiltonian is going to be replaced by the rescaled Rindler Hamiltonian $H_\text{R}$. The latter is now associated with the stretched horizon. For the total physical Hamiltonian, a similar change applies. The energy levels of the states are chosen according to the total energy perceived by fiducial observers at spatial infinity who run their clocks in Schwarzschild time $t$. Moreover, note that the rescaled value of the Rindler energy $E_\text{R}$ will contribute to the total energy.

We assume that the states assigned to the far horizon region $\specChar{out}$ satisfy $| n \rangle_\specChar{out} = | n \rangle_\specChar{I}$. In other words, we have equipped the state \eqref{eq:qbh-state} with some additional quantum structure associated with the stretched horizon. 
We are interested in the expectation value $\langle \mathcal{O} \rangle = \mathrm{tr}(\mathcal{O} \rho^\text{qbh}_\specChar{out})$ of a local operator $\mathcal{O}$ representing an observable in the far horizon region. Tracing out the near horizon states, i.e. $\rho_\specChar{out}^\text{qbh} := \mathrm{tr}_\specChar{sh} (\rho^\text{qbh}_\specChar{I})$, where $\rho^\text{qbh}_\specChar{I} = | \Psi \rangle^\text{qbh}_\specChar{I}\ {}_\specChar{I}^\text{qbh}\langle \Psi |$, reproduces, of course, the usual proper density matrix \eqref{eq:thermal-density-matrix},
\begin{equation}
    \rho_\specChar{out}^\text{qbh} = 
    \rho_\specChar{I}^\text{semi},
\label{eq:qbh-density-matrix}
\end{equation}
and thus the standard thermal value $\langle \mathcal{O} \rangle = \mathrm{tr}(\mathcal{O} \rho^\text{semi}_\specChar{I})$.

Anyway, we are still not where we want to be. The partial trace operation described above is somewhat unjustified. In the standard scenario, all states of the region $\specChar{III}$ are allowed to be traced out as the entire region is causally disconnected. Its associated states do not affect the states of $\specChar{I}$ as both sides cannot interact with each other, although they are maximally entangled. This is apparently not the case for the state introduced in \eqref{eq:qbh-state}.

In view of this shortcoming, we need a mechanism that allows us to perform an analogous partial trace within the identification picture. Therefore, an understanding of how the semiclassical limit can be consistently reproduced has not yet been obtained. This will be the content of the next section. We wish to add that a state of the form \eqref{eq:qbh-state} already eliminates an ambiguity with respect to the causal boundedness of the physical exterior.

\section*{Separability and protected indistinguishability}
\label{sec:gen-tfd-universe}
To reflect on a more appropriate decoupling scenario, let us first introduce the following eigenbasis state 
\begin{equation}
    | n \rangle_{\specChar{sh},\vartheta_n} := e^{i \vartheta_n} | n \rangle_\specChar{sh}.
\label{eq:psh-state}
\end{equation}
The phases $\vartheta_n$ shall act as random variables. Let us now consider the following generalised thermofield double state
\begin{equation}
    | \Psi \rangle^{\text{pqbh}}_\specChar{I} = \frac{1}{\sqrt{Z}} \sum_n e^{- \beta E_n / 2}\ | n \rangle_{\specChar{sh},\vartheta_n} \otimes | n \rangle_\specChar{out}.
\label{eq:pqbh-state}
\end{equation}
These states share all the properties of \eqref{eq:qbh-state}. In particular, they are invariant under the same modular Hamiltonian. 
For example, we could imagine that at $t=0$ the state \eqref{eq:qbh-state} dephases into a state that is effectively, i.e., under a set of restricted operations, described by \eqref{eq:pqbh-state}. Since randomness would ensure that the equality $(\vartheta_n - \vartheta_m) = 2 t_\vartheta (E_n - E_m)$ holds for a given pair of energy eigenvalues, \eqref{eq:pqbh-state} can be identified as \eqref{eq:qbh-state} evolved for a very long time in Schwarzschild time $t_\vartheta$, i.e., much longer than any resolution time of the restricted local operator set. Hence, local operators will not extract these topological phases as they can only be measured by operators representing nonlocal observables \cite{Verlinde:2020upt}. States of the form \eqref{eq:pqbh-state} would therefore be protected against local operations in the region near the horizon.

Provided that the phases of two distinct states are highly random, it follows that the generalised states are approximately orthogonal, i.e. ${}^\text{pqbh}_\specChar{I} \langle \Psi^\prime | \Psi \rangle^\text{pqbh}_\specChar{I} = \delta_{\vartheta \vartheta^\prime}$. Since they are protected and thus indistinguishable from \eqref{eq:qbh-state} under local operations, we may continue with the following incoherent sum 
\begin{equation}
    \rho^\text{pqbh}_\specChar{I} = \sum_{\vec \vartheta} | \Psi \rangle^{\text{pqbh}}_\specChar{I}\ {}_\specChar{I}^{\text{pqbh}} \langle \Psi |.
\end{equation}
Summing over the complete set of random phases yields a diagonal distribution $\sum e^{i(\vartheta_n - \vartheta_m)} \simeq \delta_{nm}$. The final density matrix thus takes the form of a separable state,
\begin{equation}
    \rho^\text{pqbh}_\specChar{I} = \frac{1}{Z} \sum_n e^{- \beta E_n}\ \rho^n_\specChar{sh} \otimes \rho^n_\specChar{out},
\label{eq:rho-pqbh}
\end{equation}
namely, written as a convex sum of the product states $\rho^n_\specChar{sh} = | n \rangle_\specChar{sh}\ {}_\specChar{sh} \langle n |$ and $\rho^n_\specChar{out} = | n \rangle_\specChar{out}\ {}_\specChar{out} \langle n |$ \cite{Peres:1996dw, Horodecki:1996nc}.
Interestingly, separable states as in \eqref{eq:rho-pqbh} appear in the context of topological entropy computations \cite{Wen:2016snr}. 

Consider a topologically protected system on a cylinder. Equivalently, we may imagine a two-sphere in the presence of two punctures. A Wilson line would thread from one puncture to the other and fluctuate between the different topological sectors. Cutting out the interacting centre (described by an interaction Hamiltonian and some coupling constant), with the punctures positioned to the left and right of the cuts, would then correspond to introducing two entanglement cuts, leaving behind edge state contributions on both sides. In the cylinder case, the interacting piece would be sandwiched between the left and right ends. The result would be a separable state as in \eqref{eq:rho-pqbh}, i.e., expressed as a sum over the weighted product of two density matrices associated with the left and right subsystems. For one cut a similar procedure leads to a thermal density matrix associated with the edge states at the spatial boundary, but expressed as a sum over a single weighted density matrix \cite{Qi2012general, Wen:2016snr}. A related picture has recently been discussed in the context of black hole unitarity within AdS/CFT \cite{Akal:2020ujg}.

After all, due to the effective separability, we may finally trace out the states of the stretched horizon and find the reduced density matrix for the far horizon region. We observe that the density matrix reduces to the standard semiclassical result,
\begin{equation}
    \rho_\specChar{out}^\text{pqbh} := \mathrm{tr}_\specChar{sh}(\rho^\text{pqbh}_\specChar{I}) =
    \rho_\specChar{I}^\text{semi}.
\end{equation}
Its von Neumann entropy, computed by $S_\text{vN}(\cdot) = -\mathrm{tr}(\cdot \log(\cdot))$, equals the Bekenstein-Hawking entropy, i.e.,
\begin{equation}
    S_\text{vN}(\rho^\text{pqbh}_\specChar{out}) = S_\text{vN}(\rho_\specChar{I}^\text{semi}) =
    S_\text{BH}.
\label{eq:pqbh-out-semi}
\end{equation}
Thus, topological protection of the state against local operations on the physical exterior renders the state indistinguishable from the thermal Gibbs state. This feature is reflected by \eqref{eq:pqbh-out-semi} and the following equalities
\begin{equation}
\begin{split}
    I(\rho^\text{pqbh}_\specChar{I}) 
    &=
    S_\text{vN}(\rho^\text{pqbh}_\specChar{I}) =
    S_\text{vN}(\rho^\text{pqbh}_\specChar{out}),
\end{split}
\end{equation}
where
\begin{equation}
\begin{split}
    I(\rho_\specChar{ab}) = S_\text{vN}(\rho_\specChar{a}) + S_\text{vN}(\rho_\specChar{b}) - S_\text{vN}(\rho_\specChar{ab})
\label{eq:mutual-info}
\end{split}
\end{equation}
denotes mutual information. It is a positive correlation measure and can be expressed as the relative entropy $S_\text{rel}(\rho_\specChar{ab} || \rho_\specChar{a} \otimes \rho_\specChar{b})$ between the state and the uncorrelated product state. Its value in the present case is a clear manifestation of the purely classical correlations between $\specChar{sh}$ and $\specChar{out}$. Note that $I(\rho_\specChar{I}^\text{qbh}) = 2 I(\rho_\specChar{I}^\text{pqbh})$, reflecting the presence of additional quantum correlations in the state \eqref{eq:qbh-state}. In the nonseparable case, we deduce the following inequality
\begin{equation}
    S_\text{vN}(\rho^\text{qbh}_\specChar{out}) >
    S_\text{vN}(\rho^\text{qbh}_\specChar{I}).
\end{equation}
In other words, due to the present quantum correlations, the local disorder appears larger than the global disorder, which is classically impossible \cite{Nielsen:2000zzv}. On the other hand, we observe that the separable state \eqref{eq:pqbh-state} satisfies the relation
\begin{equation}
    S_\text{vN}(\rho^\text{pqbh}_\specChar{out})  \leq
    S_\text{vN}(\rho^\text{pqbh}_\specChar{I}).
\end{equation}
In view of this quantitative difference, we conclude that the quantum black hole effectively suppresses local disorder relative to global disorder, as its protection prevents local operations in the exterior from accessing or perturbing the near-horizon structure.

\section*{Conclusions}
\label{sec:conc}
The theory of general relativity appears to contain the essential principles needed to define a consistent and well-posed Hilbert space describing the quantum black hole. 
Within this framework, the usual notion of an interior does not appear. It emerges only in the semiclassical limit, where the standard thermal Gibbs state is recovered. Gravity is meticulously sensitive and selective here; treating the curvature singularity as physically real can be interpreted as gravity’s warning that there is no underlying Hilbert space.
Of course, one may attempt to incorporate all degrees of freedom in the maximally extended geometry. Yet, doing so inevitably introduces ambiguities in the underlying causal boundedness, often manifesting as parallel universes, wormholes, and whatever it may be.

Despite an apparent departure from the well-known thermodynamic values, we have discussed that a nontrivial quantum structure in the exterior makes the identification procedure consistent with the Hawking temperature. In particular, we have seen that (topological) state protection against local operations in the physical exterior brings together the robustness of relativistic field theory and black hole unitarity. It will be compelling to further examine the relationship between the near-horizon region and the reduction prescription that removes spacetime redundancy once the horizon is rendered harmless.

Recently, it has been proposed that this protected structure emerges only at the effective level \cite{Akal:2022ssl}. We hope that the present investigations will shed light on the underlying principles. However, we also forecast that the reader may quickly be inclined to question the relevance of this perspective, given that quantum theory is regarded as the framework for formulating physics at the fundamental level.

In this context, we do not think that any sort of modification of conventional quantum mechanics could serve as a viable path forward, let alone any interpretational attempt. The framework of quantum theory does not seem to be tailored for interpretational resolutions. What would be needed is something new. In fact, no known principle rules out the possibility of a breakdown in an extremely high complexity regime. This regime lies far beyond the domains in which particle physicists successfully apply the machinery of effective field theory. However, it is precisely where the horizon region of the quantum black hole may take over.

\begin{figure}[h!]
    \centering
    \includegraphics[width=.3\linewidth]{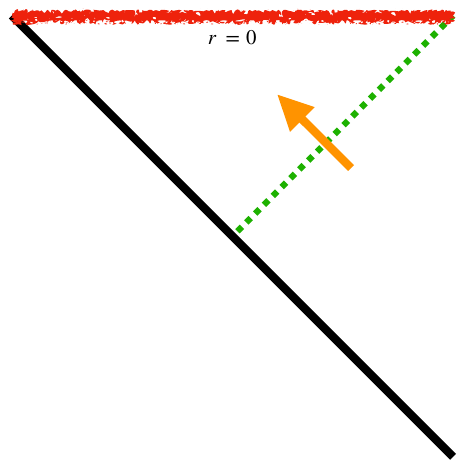}\qquad\qquad
    \includegraphics[width=.3\linewidth]{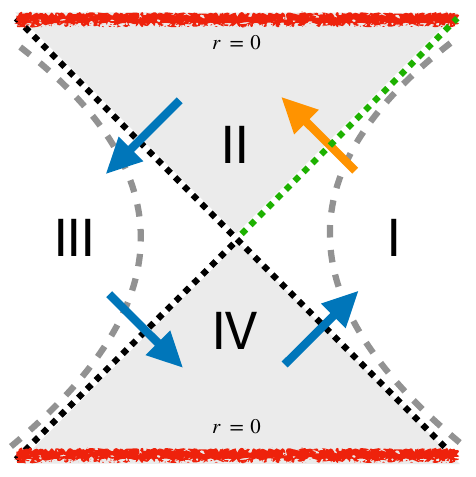}
    \caption{Penrose diagrams for the extended Schwarzschild (left) and Fronsdal-Kruskal-Szekeres coordinates (right). Killing horizons (dotted lines), i.e., null hypersurfaces to which the corresponding Killing vector fields are tangent, are traversed by null arrows. The hyperbolic curve (dashed lines) in region $\specChar{I}$ represents a fictitious cut between $\specChar{sh}$ and $\specChar{out}$.}
    \label{fig:flows}
\end{figure}

Finally, let us reconsider the Bekenstein-Hawking formula $S_\text{BH} = A(4 l_\text{P}^2)$ in light of the present discussion. In addition to the surface area of the horizon, $A$, whose role was profoundly clarified by Bekenstein, there remain two additional elements in the entropy formula that still lack a transparent understanding, leaving room for various interpretations. The first is the square of the Planck length $l_\text{P}^2 = A / (S 4)$. Admittedly, its appearance is not particularly puzzling as it underscores the importance of degrees of freedom at the fundamental scale. In essence, it reflects a geometric constraint by partitioning the surface area $A$. Qualitatively, the situation is likely similar in the transverse direction. It is therefore reasonable to assume a fundamental lower bound \eqref{eq:min-proper-distance-sh} for the proper distance from the true horizon, which in turn suggests the relation $l_\text{P}\ \hat{=}\ \Delta_\text{sh}$. However, a more subtle situation emerges with respect to the factor of $4=A/(S l_\text{P}^2)$ in the denominator. We are unaware of any established consensus on the interpretation of this value. At present, we believe that it finds a natural explanation once one considers the identification performed in the maximally extended geometry. In particular, it may arise from a rescaling effect that occurs when the four transparent Killing horizons are effectively reduced to a single horizon after eliminating the redundant copy of the physical universe; cf. figure \ref{fig:flows}. We are therefore inclined to conclude with a rather cryptic expression of the form $4\ \hat{=}\ (\specChar{III} \leftrightarrow \specChar{I})\ \backslash\ \specChar{sh}$,
which encapsulates the identification discussed throughout this work.


\bibliographystyle{JHEP}
\bibliography{BQMGH}

\end{document}